\documentclass[aps,twocolumn]{revtex4}
\usepackage{graphicx}
\usepackage{amsmath}
\usepackage{amssymb}
\usepackage{microtype}
\usepackage{soul}
\usepackage[usenames,dvipsnames,svgnames,table]{xcolor}

\begin{document}

\title{A strongly interacting gas of two-electron fermions at an orbital Feshbach resonance}

\author{G. Pagano$^{1}$, M. Mancini$^{1}$, G. Cappellini$^{2}$, L. Livi$^{2}$, C. Sias$^{3,2}$, J. Catani$^{4,2}$, M. Inguscio$^{3,1,2}$, L. Fallani$^{1,2,5}$}

\affiliation{
$^1$Department of Physics and Astronomy, University of Florence, Italy\\
$^2$LENS European Laboratory for Nonlinear Spectroscopy, Firenze, Italy\\
$^3$INRIM Istituto Nazionale di Ricerca Metrologica, Torino, Italy\\
$^4$INO-CNR Istituto Nazionale di Ottica del CNR, Sezione di Sesto Fiorentino, Italy\\
$^5$INFN Istituto Nazionale di Fisica Nucleare, Sezione di Firenze, Italy
}

\begin{abstract}
We report on the experimental observation of a strongly interacting gas of ultracold two-electron fermions with orbital degree of freedom and magnetically tunable interactions. This realization has been enabled by the demonstration of a novel kind of Feshbach resonance occurring in the scattering of two $^{173}$Yb atoms in different nuclear and electronic states. The strongly interacting regime at resonance is evidenced by the observation of anisotropic hydrodynamic expansion of the two-orbital Fermi gas. These results pave the way towards the realization of new quantum states of matter with strongly correlated fermions with orbital degree of freedom.
\end{abstract}

\maketitle

Recent theoretical and experimental work has evidenced how quantum gases of two-electron atoms represent extraordinary systems for the development of a new generation of quantum simulators of fermionic matter. They provide realizations of multicomponent fermionic gases with tunable number of nuclear spin states and SU(N)-symmetric interactions \cite{taie2012,zhang2014,pagano2014}. In addition, they offer the unique feature of an additional electronic (orbital) degree of freedom, over which coherent quantum control can be achieved by means of the technology developed in the context of optical atomic clocks \cite{ludlow2015}. Recently, a strong spin-exchange interaction between $^{173}$Yb fermions in different orbital states has been observed, paving the way to the realization of orbital quantum magnetism and of the Kondo lattice model \cite{scazza2014,cappellini2014}. Despite these exciting perspectives, up to now two-electron atoms lacked the tunability of interactions that is provided by Feshbach resonances in the case of alkalis. Indeed, magnetic Feshbach resonances in ultracold gases of alkali atoms \cite{chin2010} have allowed breakthrough achievements, including unprecedented studies of strongly interacting fermions, with the demonstration of high-density molecular gases and the exploration of fermionic superfluidity at the BEC-BCS crossover \cite{inguscio2007}. A similar tunability for two-electron atoms would open totally new avenues, but the zero electronic angular momentum in their ground state precludes the existence of accessible and useful magnetic Feshbach resonances. While optical Feshbach resonances in two-electron atoms have been proposed \cite{ciurylo2005} and observed experimentally \cite{blatt2011,enomoto2008}, their actual implementation still suffers from severe intrinsic difficulties, such as heating and losses, preventing the observation of true many-body quantum physics, although novel promising schemes have been very recently investigated \cite{taie2015}.

In this work we report on the first realization of a strongly-interacting gas of two-electron fermionic atoms with orbital degree of freedom and tunable interactions. We take advantage of a recently-proposed {\it orbital Feshbach resonance}  \cite{zhang2015} affecting the scattering between $^{173}$Yb atoms in different electronic states. We observe a hydrodynamic expansion of the Fermi gas in the strongly interacting regime and use it to identify the resonance position. We also verify the predicted scaling of the resonance centers for different spin mixtures, arising from the SU(N) symmetry of $^{173}$Yb Fermi gases, and characterize the atom losses across the resonance, showing lifetimes that could allow for the investigation of the BEC-BCS crossover in two-orbital binary mixtures.

\begin{figure}[t!]
\begin{center}
\includegraphics[width=\columnwidth]{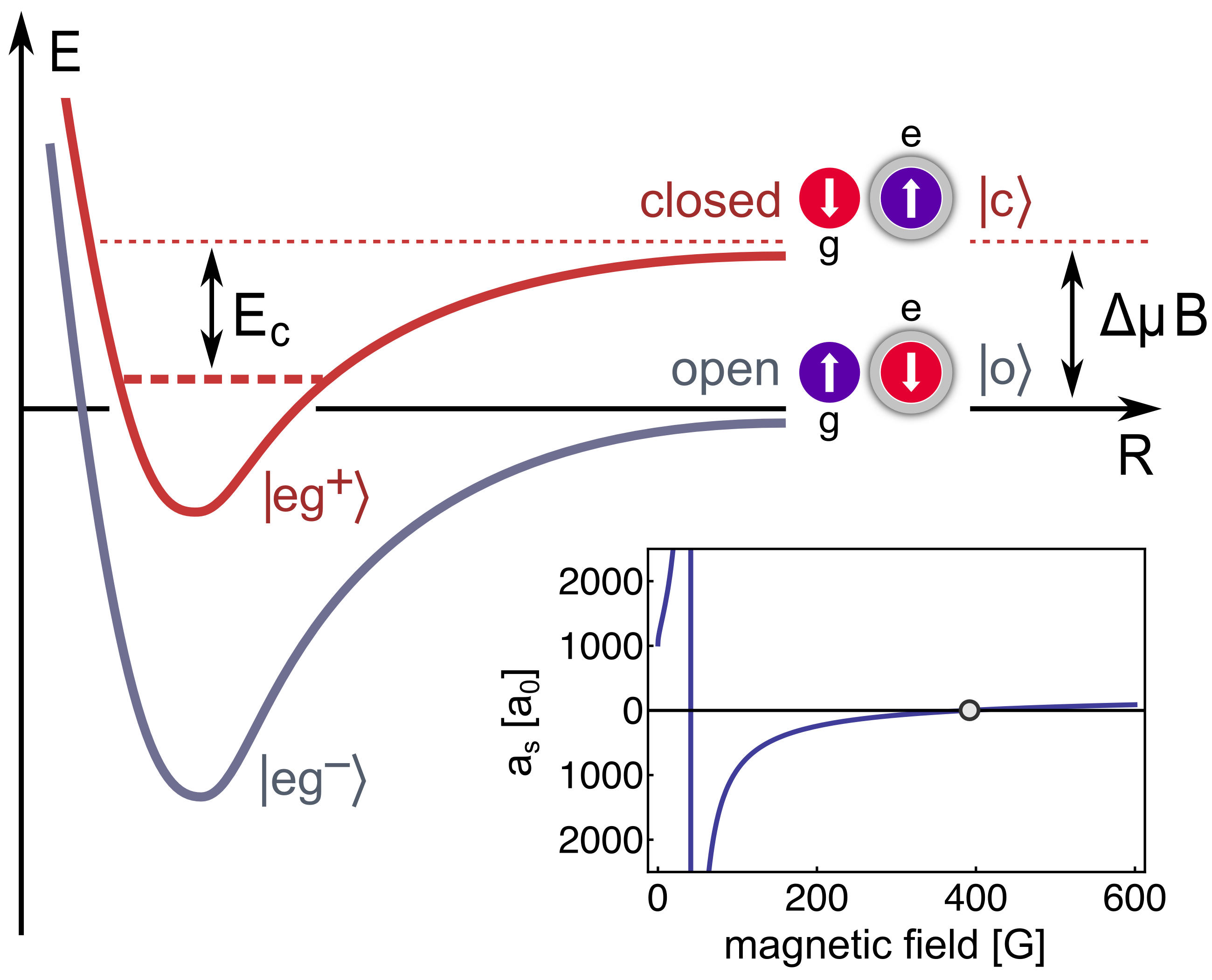}
\end{center}
\caption{Scheme of the $^{173}$Yb orbital Feshbach resonance affecting the scattering between one $\left| g \right>$ $=$ $^1S_0$ atom and one $\left| e \right>$ $=$ $^3P_0$ atom in two different spin states (see text for details). The inset shows the behavior of the scattering length for $\Delta m=5$ as modeled in Ref. \cite{zhang2015}, with the center of the resonance adjusted to match our experimental observations in Fig. 3 (the circle identifies the predicted position of the zero crossing).}
\label{fig1}
\end{figure}

The physical idea behind an orbital Feshbach resonance (OrbFR) is similar to that of a magnetic Feshbach resonance, although with some distinctive differences.  We consider two $^{173}$Yb atoms in two different electronic (orbital) states $\left| g \right>$ $=$ $^1S_0$ (stable) and $\left| e \right>$ $=$ $^3P_0$ (metastable, $\sim 20$ s lifetime), and different nuclear spin states $\left| \uparrow \right>$ and $\left| \downarrow \right>$. When the atoms are separated, interactions between them are negligible and the relevant two-body eigenstates are $\left| o \right> = \left| g\uparrow ; e\downarrow \right>$ and $\left| c \right> = \left| g\downarrow ; e\uparrow \right>$, that we name as {\it open} and {\it closed} collisional channels, respectively (see Fig. \ref{fig1}). The energy separation between the two channels is given by the differential Zeeman shift $\Delta \mu B = \delta_g \mu_N \Delta m B$ (where $\delta_g$ is the differential Land\'e factor between $\left| e \right>$ and $\left| g \right>$, and $\mu_N$ is the nuclear magneton). In $^{173}$Yb $\Delta \mu = 113$ Hz/G $\times$ $\Delta m$, where $\Delta m$ is the difference between the spin quantum numbers $m_\uparrow$ and $m_\downarrow$. As the interatomic distance decreases, the appropriate basis for the description of the scattering is given by the orbital symmetric $\left| eg^+ \right> = (\left| c \right> - \left| o \right>)/\sqrt{2}$ and antisymmetric states $\left| eg^- \right> = (\left| c \right> + \left| o \right>)/\sqrt{2}$, which are associated to two distinct molecular potentials, giving rise to two very different scattering lengths $a_{eg}^+$ and $a_{eg}^-$, respectively \cite{cappellini2014,scazza2014}. The difference between them determines an effective coupling between open and closed channels, which becomes resonant when the differential Zeeman energy $\Delta \mu B$ equals the binding energy $E_c$ of the least bound state in the closed channel. Luckily, the orbital-symmetric potential features a shallow bound state with $|E_c|$ in the tens of kHz range, which corresponds to a large background scattering length $a_{eg}^+$. Thanks to this peculiar molecular potential of $^{173}$Yb, it is possible to use very convenient fields of the order of only tens of Gauss to magnetically tune the closed-channel bound state to be resonant with the open channel, despite the extremely weak sensitivity of two-electron atoms to external magnetic fields (about $10^3$ times smaller than in alkalis).

An ultracold gas of $N \simeq 6\times10^4$ fermionic $^{173}$Yb atoms is prepared in a crossed optical dipole trap, operating at the magic wavelength $\lambda_m = 759$ nm in order not to shift the ultranarrow $^1S_0$ $\rightarrow$ $^3P_0$ clock transition used for the preparation of the initial state. One of the trap beams is tightly focused, resulting in a cigar-shaped trap with angular frequencies $\vec{\omega}=(\omega_x,\omega_y,\omega_z)=2\pi \times (13,188,138)$ Hz. By proper spin manipulation protocols (relying on optical pumping and spin-selective removal of atoms by resonant excitation on the $^1S_0$ $\rightarrow$ $^3P_1$ transition), the gas is initially prepared in a balanced spin mixture of $N/2 + N/2$ ground-state atoms in nuclear spin states $m_\uparrow$ and $m_\downarrow$, with different $\Delta m = m_\uparrow - m_\downarrow$. The temperature of the Fermi gas is $T \simeq 0.15\, T_F$ (where $T_F$ is the Fermi temperature) and the peak density is $n \simeq 2.4 \times 10^{13}$ cm$^{-3}$ per spin component. In order to excite the atoms to the $\left| e \right>$ state, we first slowly turn on the intensity of a 1D optical lattice at the magic wavelength with depth $30 E_R$ (where $E_R=h^2/2m\lambda_m^2$ is the recoil energy, with $m$ the atomic mass). Along the direction of the lattice, we shine a 400 $\mu$s resonant pulse of light on the 578 nm clock transition $^1S_0$ $\rightarrow$ $^3P_0$ \cite{cappellini2015}, in such a way as to perform the optical excitation in the Lamb--Dicke regime \cite{ludlow2015}. This allows us to excite the atoms to the $\left| e \right>$ state with an efficiency $\gtrsim 80\%$ without imparting an optical momentum kick, which is important to maintain the system at equilibrium. The excitation is performed at a large magnetic field $B_{exc}$ ranging from 72 to 167 G, in order to clearly resolve the Zeeman structure of the transition and excite only one spin state. In this way we can selectively access the open or the closed channel (see Fig. \ref{fig1}). After the excitation, we adiabatically switch off the optical lattice in 100 ms to recover a 3D atomic gas. Then, we change the magnetic field intensity to the desired value $B$ in about 2 ms. The trap is suddenly switched off and, after a time of flight $t_{TOF}$, the atoms remaining in the ground state $\left|g\right>$ are imaged with a resonant pulse of light along $\hat{z}$ on the $^1S_0$ $\rightarrow$ $^1P_1$ transition. During the first 5 ms of the expansion the magnetic field is kept at the $B$ value, thus allowing the atoms to release their interaction energy into kinetic energy.

\begin{figure}[t!]
\begin{center}
\includegraphics[width=\columnwidth]{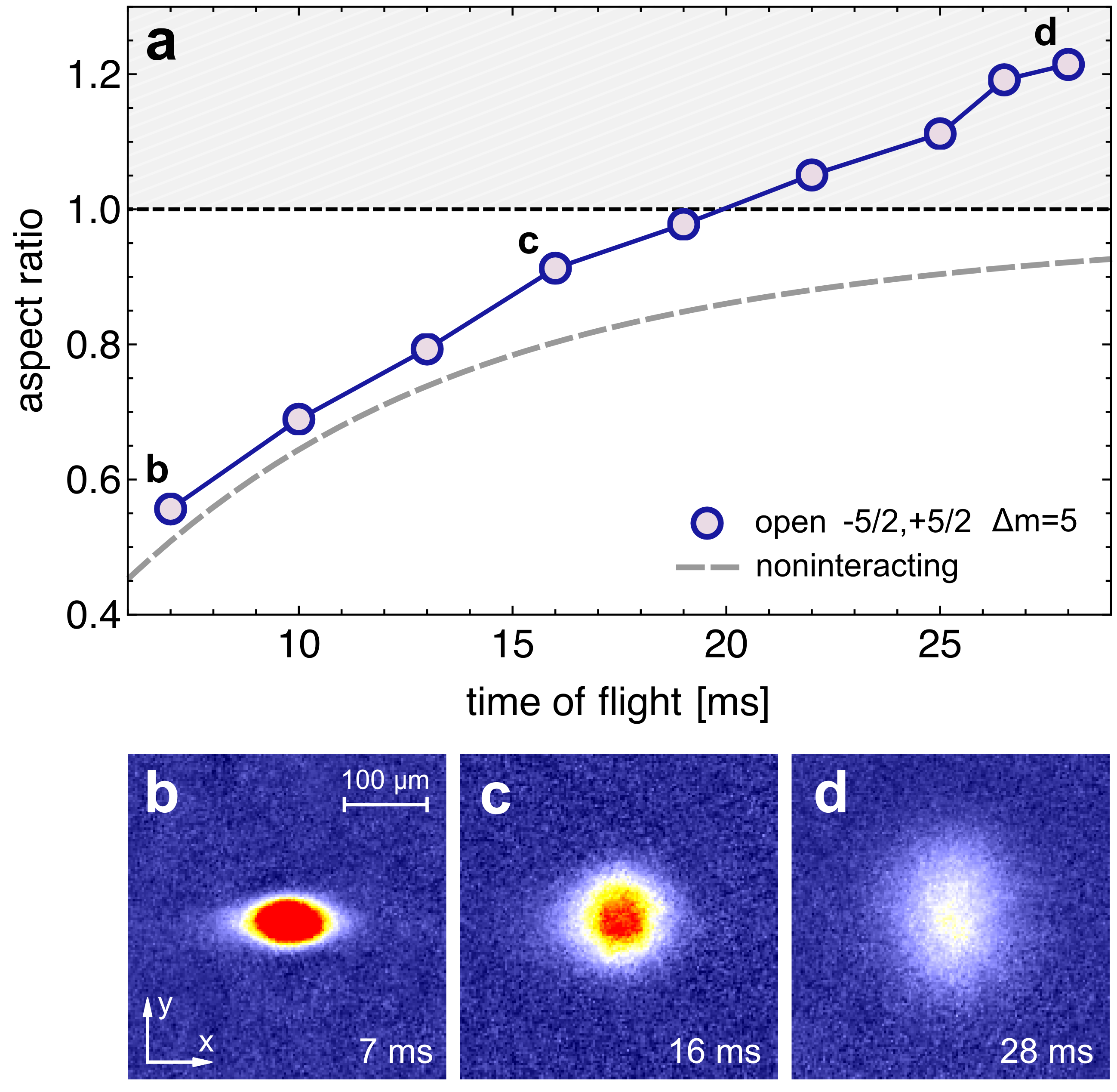}
\end{center}
\caption{Anisotropic expansion of a strongly interacting $^{173}$Yb Fermi gas, initially prepared in the open channel of a spin mixture with $\Delta m =5$, at a magnetic field $B=41$ G. {\bf a.} The circles show the aspect ratio of the expanded $\left| g \right>$ atomic cloud as a function of the time of flight. The long-dashed grey line shows the behavior expected for an ideal Fermi gas. The short-dashed black line indicates the unit limit above which the experimental aspect ratio gets inverted because of hydrodynamic expansion. {\bf b-d.} False-color absorption images of the $\left| g \right>$ component of the atomic cloud after excitation in the open channel for increasing time of flight ({\bf b.} 7 ms {\bf c.} 16 ms {\bf d.} 28 ms). }
\label{fig2}
\end{figure}

Fig. \ref{fig2}a shows the evolution of the aspect ratio of the Fermi gas after the release from the trap as a function of the time of flight. The gas is prepared in the open channel of a spin mixture $(m_\downarrow=-5/2,m_\uparrow=+5/2)$ with $\Delta m=5$. The aspect ratio is defined as the ratio $R_y/R_x$ of the expanded atomic cloud size along $\hat{y}$ to the size along $\hat{x}$. In the case of a noninteracting Fermi gas (long-dashed lines) the expansion is ballistic, eventually resulting in a spherical shape and in an aspect ratio value of 1 for sufficiently long expansion times (much larger than the inverse trap frequencies). The experimental circles, showing the behavior of the interacting spin mixture at a magnetic field $B=41$ G, clearly show an inversion of the cloud shape from prolate to oblate, with an aspect ratio exceeding 1 after a $t_{TOF}\simeq 18$ ms.
Figs. \ref{fig2}b-d show false-color absorption images of the atomic cloud for different times of flight, as specified in Fig. \ref{fig2}a. The observation of the aspect ratio inversion is a hallmark of hydrodynamic expansion of the Fermi gas, which occurs in the regime of strong interactions, as observed for alkali fermionic gases close to magnetic Feshbach resonances \cite{ohara2002}. 
In the hydrodynamic limit the collisional rate $\Gamma$ is larger than the geometric trapping frequency $\bar{\omega}=(\omega_x \omega_y \omega_z)^{1/3}$, causing a faster expansion along the tightly confined axis of the harmonic trap because of the larger density gradient.
The observation of the aspect ratio inversion is an unambiguous signature of a collisional hydrodynamic regime and of strong interactions, which is a necessary condition for the onset of fermionic superfluidity \cite{greiner2003,jochim2003,zwierlein2003}. The quantitative value of the aspect ratio at resonance is lower than that predicted by the hydrodynamic equations of superfluids at resonant interactions ($\approx 2$ at our maximum expansion time) \cite{menotti2002}. This discrepancy has been already observed in \cite{regal2003}, and could be ascribed to the expanded gas not being fully in the hydrodynamic regime, possibly due to the narrow character of the OrbFR \cite{hazlett2012}.

\begin{figure}[t!]
\begin{center}
\includegraphics[width=\columnwidth]{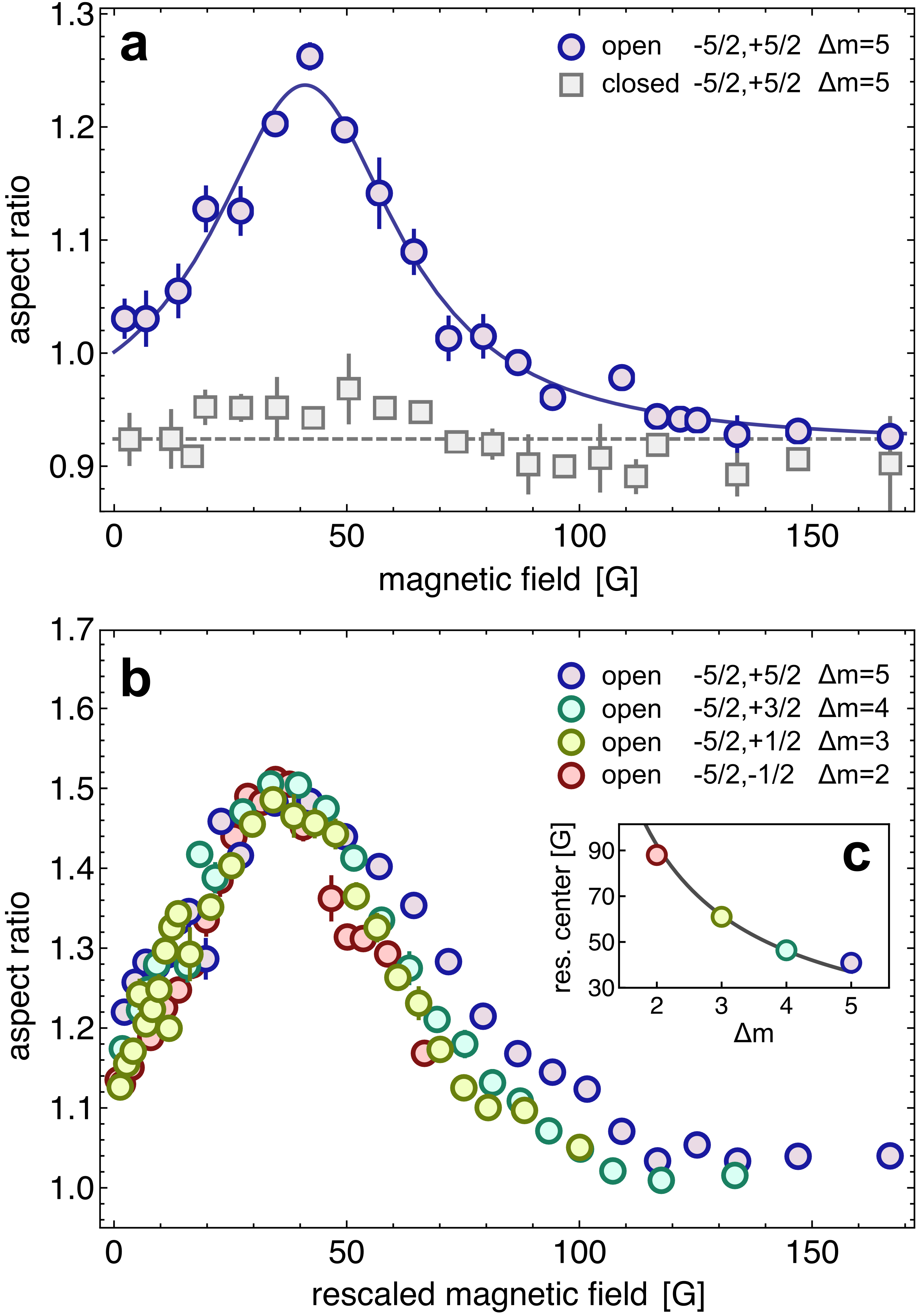}
\end{center}
\caption{Aspect ratio of the atomic cloud after a time of flight $t_{TOF}=28$ ms. {\bf a.} Comparison between expansion in the open channel and expansion in the closed channel for a mixture with $\Delta m=5$. The solid line is a fit with a Lorentzian function. The aspect ratio background signal is less than unity because the expansion is not fully in the far-field limit. {\bf b} Comparison between open-channel mixtures with different $\Delta m$, plotted as a fuction of a rescaled magnetic field $\bar{B}=B \Delta m/5$. Different colors refer to different spin states, as explained in the legend. The collapse of the different datasets onto the same curve is a verification of the OrbFR scaling law, which is in turn a direct consequence of the SU(N) symmetry of two-electron atoms. For these measurements we have used a different excitation scheme (the atoms are excited directly in the 3D trap without the applicaton of the optical lattice) and a different trap geometry $\vec{\omega}=2\pi \times (22,181,139)$ Hz, which cause the peak aspect ratio to be different from that of the data in panel {\bf a}. {\bf c.} Resonance centers of the data shown in {\bf b} (before rescaling the magnetic field) plotted versus $\Delta m$. The line is a fit with the expected $\Delta m^{-1}$ behavior.
}
\label{fig3}
\end{figure}

We observe that the asymptotic value of the aspect ratio in the open channel depends on the magnetic field value at which the expansion takes place, which is a strong evidence for the existence of the OrbFR predicted in Ref. \cite{zhang2015}. Indeed, the anisotropy can be used as a tool to characterize the Feshbach resonance, as already studied with alkali atomic gases \cite{regal2003}. In Fig. \ref{fig3}a we show the difference in the aspect ratio between the open channel and the closed channel for a spin mixture with $\Delta m=5$ as a function of the magnetic field at $t_{TOF}=28$ ms. In the open channel a clear resonant behavior is observed, with a maximum that is located at a magnetic field $B=(41 \pm 1 )$ G, signalling the enhancement of the elastic collisional rate at the Feshbach resonance. While we exclude possible confinement-induced shifts because of the low trapping frequencies in the 3D trap geometry, finite-energy effects could affect the resonance position since $ | E_c | \sim h \times 20$ kHz is only about one order of magnitude larger than the temperature and the Fermi energy $k_B T_F \sim h \times 4$ kHz of the gas \cite{zhang2015} (the latter corresponds to a magnetic field $B=k_B T_F/\Delta \mu \simeq 7$ G for $\Delta m=5$, which might also contribute to the width of the observed features). Nevertheless, from the fitted value of the resonance center, using Eq. (13) of Ref. \cite{zhang2015}, we can extract $a_{eg}^+ + a_{eg}^- \approx 2100 a_0$.  However, we must note that this is only a rough estimate, given the narrow character of the observed resonance \cite{chin2010}.

A more precise value would require further theoretical and experimental investigation, that will be postponed to future works. The value of $a_{eg}^+ + a_{eg}^-$ also allows for an estimation of the magnetic field value at which the zero crossing of the scattering length occurs. Including it in Eq. (12) of Ref. \cite{zhang2015}, we obtain the behavior of the scattering length for $\Delta m = 5$, which is plotted in the inset of Fig. \ref{fig1}. The curve shows a zero crossing point around 400 G, which lays outside the range of our experimental setup (170 G).

We have repeated the measurements for smaller $\Delta m$, evidencing similar resonances at significantly larger magnetic fields, as shown by the resonance centers plotted in Fig. \ref{fig3}c.
The data are plotted in Fig. \ref{fig3}b versus a rescaled magnetic field $\bar{B}=B \Delta m/5$. The different datasets clearly show a very similar dependence on $\bar{B}$, which is a distinctive feature of the OrbFR mechanism. As a matter of fact, the magnetic field at which the resonance is located is expected to scale as $\Delta m^{-1}$ \cite{zhang2015}. This scaling law directly follows from the SU(N) invariance of scattering in $^{173}$Yb, which determines $E_c$ to be independent of $\Delta m$. As a consequence, the Zeeman energy $\delta_g \mu_N \Delta m B$ for which the coupling between the scattering channels is resonant, has to be the same for different $\Delta m$, from which the scaling law $B \sim \Delta m^{-1}$ follows. The collapse of the experimental data onto the same curve is, therefore, a verification of the OrbFR scaling law, which is in turn a strong evidence for the SU(N) symmetry of two-electron atoms. 

\begin{figure}[t!]
\begin{center}
\includegraphics[width=\columnwidth]{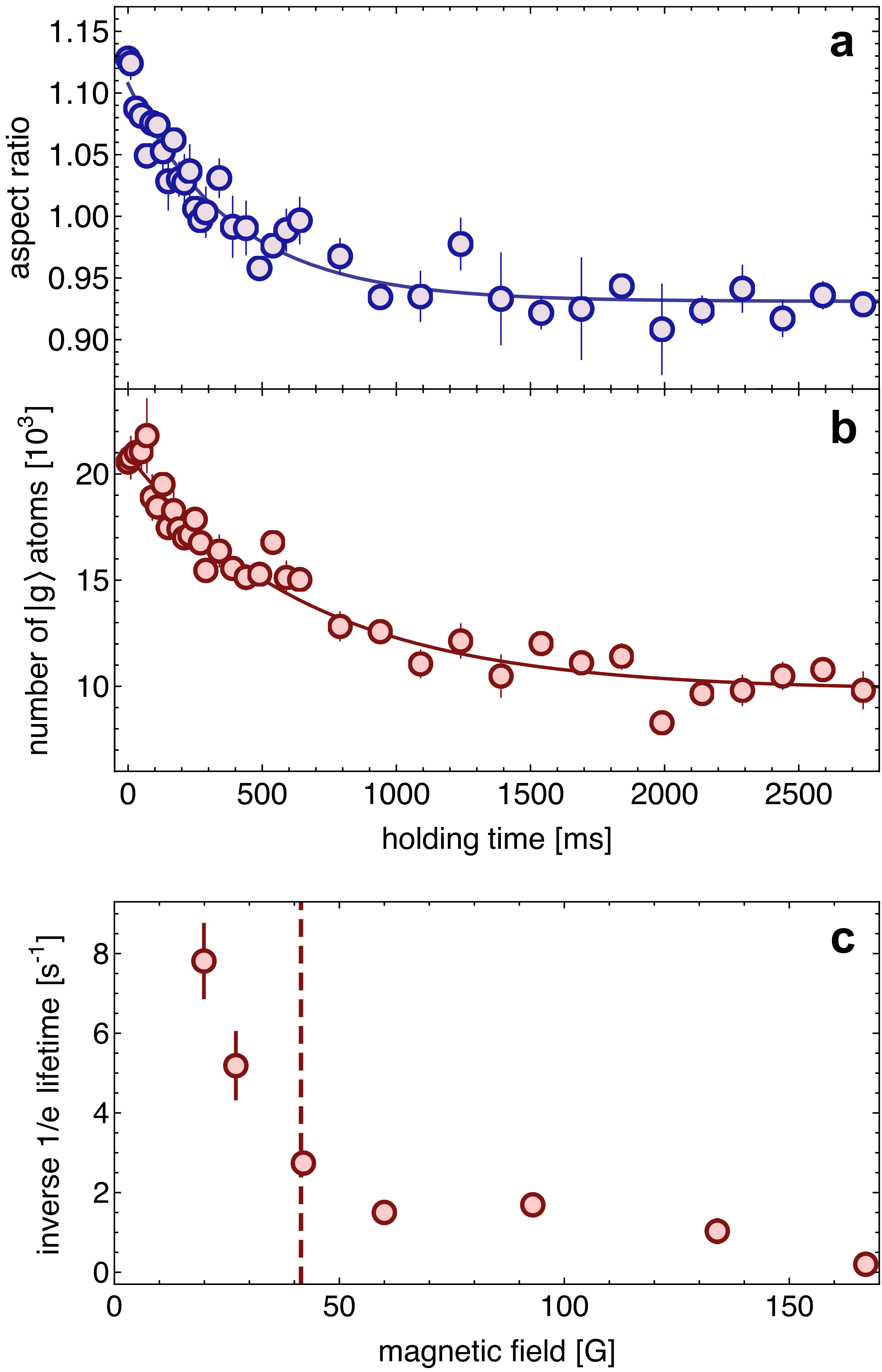}
\end{center}
\caption{{\bf a.} Evolution of the aspect ratio for a $\Delta m=5$ cloud, initially prepared in the open channel, after different holding times in the trap at $B=60$ G and a fixed time of flight $t_{TOF}=28$ ms. {\bf b.} Number of atoms for the same experimental parameters as in {\bf a.} The lines in {\bf a} and {\bf b} are exponential fits. {\bf c.} Inverse $1/e$ lifetime of the $\Delta m=5$ open-channel mixture as a function of the magnetic field. }
\label{fig4}
\end{figure}

For future studies of many-body physics in this novel experimental system, it is important to assess the effect of inelastic collisions at the Feshbach resonance. To this aim, we investigate the two-orbital Fermi gas, initially prepared in the open channel, for different holding times $t_{hold}$ in the trap. Fig. \ref{fig4}a shows the measured aspect ratio as a function of the holding time at $B=60$ G in the trap, while Fig. \ref{fig4}b shows the number of atoms $N_g$ remaining in the $\left| g \right>$ state. From these measurements one can infer several important conclusions. First, we observe that the aspect ratio exhibits a slow decay (with $1/e$ lifetime $\tau \simeq 380$ ms) towards the value (slightly smaller than 1) expected for a weakly interacting Fermi gas for the finite time-of-flight $t_{TOF}= 28$ ms. On the same timescale, $N_g$ is decreasing towards a nonzero value. This behavior can be interpreted as the result of inelastic collisions between $\left| e \right>$ and $\left| g \right>$ atoms, which empty the ground state and cause the two-orbital Fermi gas to become progressively less interacting and to abandon the conditions for being collisionally hydrodynamic.  The asymptotic nonzero value of $N_g$ in Fig. \ref{fig4}b can be interpreted as an excess of $\left| g \right>$ atoms, caused by imperfections in the preparation of the initial state, which remain in the trap after the $\left| e \right>$ atoms have been lost because of inelastic collisions. 

We note that the smooth decay of the aspect ratio in Fig. \ref{fig4}a is an evidence of the absence of shape excitations of the atomic cloud (which should have occurred on timescales of the order of the inverse trap frequencies), which confirms the adiabaticity of the excitation procedure. We have also verified that in the 3D geometry employed in the experiment the inter-orbital spin-exchange dynamics observed in Ref. \cite{cappellini2014} does not take place because of the small spin-exchange interaction energy $V_{ex}$: the asymptotic closed and open channels are well defined by the differential Zeeman energy $\Delta\mu B \gg V_{ex}$ and close to the Feshbach resonance we do not observe any significant repopulation of atoms in the $\left| g \downarrow \right>$ state of the closed channel.

Finally, we investigate the lifetime of the Fermi gas as a function of the magnetic field across the OrbFR, as illustrated in Fig. \ref{fig4}c. The lifetimes are estimated from a fit of the experimental data with a single exponential function (see e.g. the line in Fig. \ref{fig4}b), used to globally quantify the losses without differentiating among the possible decay processes, which include atom+dimer and dimer+dimer inelastic collisions \cite{Petrov2005, Ho2011}. We observe a strongly asymmetric lifetime with respect to the center of the Feshbach resonance, with an increase in the loss rate on the BEC side, as expected by the activation of the aforementioned inelastic channels. The presence of multiple decay channels cause the single exponential fit not to be fully satisfactory in the description of the data, especially on the BEC side of the resonance, and further investigation is needed in order to ascertain the fundamental processes at the basis of the atom losses. Nevertheless, the observed lifetime at the resonance seems to be rather long, with a $1/e$ lifetime of $\approx 350$ ms measured at $B= 41$ G, making this OrbFR a promising tool for future investigations of many-body physics by using two-electron atoms.

In conclusion, we have reported on the first realization of a gas of strongly-interacting two-orbital fermions with resonant interactions. This observation is the starting point for a whole new range of experimental investigations, ranging from the investigation of the BEC-BCS crossover in an ultracold gas of fermions with orbital degree of freedom, including the realization of novel forms of topological superfluids with spin-orbit coupling \cite{zhang2015}, to the investigation of two-orbital Hubbard models \cite{gorshkov2010} with tunable interactions. 

We acknowledge inspiring discussions with H. Zhai, S. Stringari, F. Minardi and M. Zaccanti. This work has been supported by EU FP7 SIQS, MIUR PRIN2012 AQUASIM, INFN FISh.

\end{document}